\documentclass[twocolumn,showpacs,preprintnumbers,amsmath,amssymb]{revtex4}
\usepackage{dcolumn}
\usepackage{bm}

\begin {document}

\title {The strong superadditivity conjecture holds for the quantum depolarizing channel in any dimension}

\author {Grigori G. Amosov}

\email {gramos@mail.ru}

\affiliation {Department of Higher Mathematics\\
Moscow Institute of Physics and Technology\\ Dolgoprudny
141700\\RUSSIA}

\date{\today}

\begin {abstract}
Given a quantum channel $\Phi $ in a Hilbert space $H$ put $\hat
H_{\Phi}(\rho)=\min \limits _{\rho _{av}=\rho }\Sigma
_{j=1}^{k}\pi _{j}S(\Phi (\rho _{j}))$, where $\rho _{av}=\Sigma
_{j=1}^{k}\pi _{j}\rho _{j}$, the minimum is taken over all
probability distributions $\pi =\{\pi _{j}\}$ and states $\rho
_{j}$ in $H$, $S(\rho)=-Tr\rho\log\rho$ is the von Neumann
entropy of a state $\rho$. The strong superadditivity conjecture
states that $\hat H_{\Phi \otimes \Psi}(\rho)\ge \hat
H_{\Phi}(Tr_{K}(\rho))+\hat H_{\Psi}(Tr_{H}(\rho))$ for two
channels $\Phi $ and $\Psi $ in Hilbert spaces $H$ and $K$,
respectively. We have proved the strong superadditivity
conjecture for the quantum depolarizing channel in any dimensions.

\end {abstract}


\maketitle

\section {Introduction}

A linear trace-preserving map $\Phi $ on the set of states
(positive unit-trace operators) $\mathfrak{S}(H)$ in a Hilbert
space $H$ is said to be a quantum channel if $\Phi ^{*}$ is
completely positive (\cite {Hol}). The channel $\Phi $ is called
bistochastic if $\Phi (\frac {1}{d}I_{H})=\frac {1}{d}I_{H}$.
Here and in the following we denote by $d$ and $I_{H}$ the
dimension of $H,\ dimH=d<+\infty ,$ and the identity operator in
$H$, respectively.

Given a quantum channel $\Phi $ in a Hilbert space $H$ put (\cite
{Sh})
\begin {equation}\label {quant}
\hat H_{\Phi}(\rho)=\min \limits _{\rho _{av}=\rho}\sum \limits
_{j=1}^{k}\pi _{j}S(\Phi (\rho _{j})),
\end {equation}
where $\rho _{av}=\sum \limits _{j=1}^{k}\pi _{j}\rho _{j}$ and
the minimum is taken over all probability distributions $\pi
=\{\pi _{j}\}$ and states $\rho _{j}\in \mathfrak{S}(H)$. Here
and in the following $S(\rho)=-Tr(\rho\log \rho)$ is the von
Neumann entropy of a state $\rho $. The strong superadditivity
conjecture states that
\begin {equation}\label {strong}
\hat H_{\Phi \otimes \Psi}(\rho)\ge \hat
H_{\Phi}(Tr_{K}(\rho))+\hat H_{\Psi}(Tr_{H}(\rho)),
\end {equation}
$\rho \in \mathfrak{S}(H\otimes K)$ for two channels $\Phi $ and
$\Psi $ in Hilbert spaces $H$ and $K$, respectively.

The infimum of the output entropy of a quantum channel $\Phi $ is
defined by the formula
\begin {equation}\label {addit}
S_{min}(\Phi)=\inf \limits _{\rho\in \mathfrak{S}(H)}S(\Phi
(\rho)).
\end {equation}
The additivity conjecture for the quantity $S_{min} (\Phi)$ states
(\cite {Hol2})
\begin {equation}\label {conj}
S_{min} (\Phi\otimes \Psi)=S_{min} (\Phi)+S_{min} (\Psi)
\end {equation}
for an arbitrary quantum channel $\Psi $. It was shown in (\cite
{Sh}) that if the strong superadditivity conjecture holds, then
the additivity conjecture for the quantity $S_{min}$ holds too.
Nevertheless the conjecture (\ref {strong}) is stronger than
(\ref {addit}).

In the present paper we shall prove the strong superadditivity
conjecture for the quantum depolarizing channel for all dimensions
of $H$.

\section {The estimation of the output entropy}

Our approach is based upon the estimate of the output entropy
proved in \cite {C02}. Combining formulae (111) and (112) in
\cite {C02} we get the lemma formulated below.

{\bf Lemma.}{\it Let $\Phi _{dep}(\rho)=(1-p)\rho+\frac
{p}{d}I_{H},\ \rho\in \mathfrak{S} (H),\ 0\le p\le \frac
{d^{2}}{d^{2}-1},$ be the quantum depolarizing channel in the
Hilbert space $H$ of the dimension $d$. Then, for any quantum
channel $\Psi $ there exist the orthonormal basis $\{e_{s},\ 1\le
s\le d\}$ in $H$ and $d$ states $\rho_{s}\in \mathfrak {S}(K),\
1\le s\le d,$ such that
\begin {equation}\label {XJ}
S((\Phi _{dep}\otimes \Psi)(\rho))\ge -(1-\frac {d-1}{d}p)\log
(1-\frac {d-1}{d}p)-
\end {equation}
$$
\frac {d-1}{d}p\log \frac {p}{d}+ \frac {1}{d}\sum \limits
_{s=1}^{d}S(\Psi (\rho_{s}))
$$
and
$$
\frac {1}{d}\sum \limits _{s=1}^{d}\rho _{s}=Tr_{H}(\rho ),
$$
where $\rho\in \mathfrak{S} (H\otimes K),\
\rho_{s}=dTr_{H}((|e_{s}><e_{s}|\otimes I_{K})\rho)\in \mathfrak
{S} (K),\ 1\le s\le d$. }

In the present paper our goal is to prove the following theorem.

{\bf Theorem.}{\it Let $\Phi _{dep}$ be the quantum depolarizing
channel in the Hilbert space of the dimension $d$. Then, for an
arbitrary quantum channel $\Psi $ in a Hilbert space $K$ the
strong superadditivity conjecture holds, i.e.
\begin {equation}\label {theorem}
\hat H_{\Phi _{dep}\otimes \Psi}(\rho)\ge \hat H_{\Phi
_{dep}}(Tr_{K}(\rho))+\hat H_{\Psi}(Tr_{H}(\rho)).
\end {equation}
}

Proof.

Suppose that
\begin {equation}\label {avr}
\rho =\sum \limits _{j=1}^{k}\pi _{j}\rho _{j}
\end {equation}
and the states $\rho _{j},\ 1\le j\le k,$ form the optimal
ensemble for (\ref {quant}) in the sense that
\begin {equation}\label {avr2}
\hat H_{\Phi _{dep}\otimes \Psi}(\rho)=\sum \limits _{j} \pi
_{j}S((\Phi _{dep}\otimes \Psi)(\rho _{j}))
\end {equation}
Applying (\ref {XJ}) to each element of the sum in (\ref {avr2})
we get
\begin {equation}\label {A1}
\hat H_{\Phi _{dep}\otimes \Psi}(\rho)\ge -(1-\frac
{d-1}{d}p)\log (1-\frac {d-1}{d}p)-
\end {equation}
$$
\frac {d-1}{d}p\log \frac {p}{d}+\frac {1}{d}\sum \limits
_{j=1}^{k}\pi _{j}\sum \limits _{s=1}^{d}S(\Psi (\rho_{js})),
$$
where $\rho_{js}=dTr_{H}((|e_{js}><e_{js}|\otimes I_{K})\rho
_{j})\in \mathfrak {S} (K),\ 1\le j\le d$, and each the set
$\{e_{js},\ 1\le s\le d\}$ forms the orthonormal basis of $H$ for
$1\le j\le k$.

It follows from Lemma that
\begin {equation}\label {Sup1}
\frac {1}{d}\sum \limits _{j=1}^{k}\pi _{j}\sum \limits
_{s=1}^{d}\Psi (\rho_{js})=\sum \limits _{j=1}^{k}\pi _{j}\Psi (
Tr_{H}(\rho _{j}))=\Psi (Tr_{H}(\rho)).
\end {equation}
The equality (\ref {Sup1}) results in
\begin {equation}\label {E2}
\frac {1}{d}\sum \limits _{j=1}^{k}\pi _{j}\sum \limits
_{s=1}^{d}S(\Psi (\rho_{js}))\ge \hat H_{\Psi}(Tr_{H}(\rho)).
\end {equation}
Notice that the quantity (\ref {quant}) is always bounded from
below by the quantity (\ref {addit}). For the quantum
depolarizing channel $\Phi _{dep}$ (\ref {quant}) coincides with
(\ref {addit}) for any state because (\ref {addit}) is achieved on
any pure input state due to the covariance property of $\Phi
_{dep}$. Thus, we get
\begin {equation}\label {E3}
\hat H_{\Phi _{dep}}(\rho)=-(1-\frac {d-1}{d}p)\log (1-\frac
{d-1}{d}p)-
\end {equation}
$$
\frac {d-1}{d}p\log \frac {p}{d}=S_{min} (\Phi _{dep})
$$
for any state $\rho \in \mathfrak {S}(H)$. Taking into account
(\ref {A1}),(\ref {E2}) and (\ref {E3}) we get
$$
\hat H_{\Phi _{dep}\otimes \Psi}(\rho)\ge \hat H_{\Phi _{dep}
}(Tr_{K}(\rho))+\hat H_{\Psi}(Tr_{H}(\rho)),
$$
$\rho \in \mathfrak{S}(H\otimes K).$ Thus, the strong
superadditivity conjecture for the quantum depolarizing channel
is proved.

$\Box $

\section {Conclusion}

At the first time the additivity conjecture (\ref {conj}) for the
quantum depolarizing channel was proved in \cite {C02}. The method
was based upon the estimation of $l_{p}$-norms of the channel. On
the other hand in the papers \cite {Amo, Amo1, Amo2} it was shown
that the decreasing property of the relative entropy also can be
used to prove the additivity conjecture for some partial cases at
least. In the present paper we have proved that the estimation of
the output entropy obtained in \cite {C02} allows to prove the
strong superadditivity conjecture (\ref {strong}) for the quantum
depolarizing channel. One of a possible basis for considering the
strong superadditivity conjecture can be drawn from the paper
\cite {Sh}. There was presented the proof of the global
equivalence of the additivity conjecture for the constrained
channels and the strong superadditivity conjecture.

\section*{Acknowledgments}

The author is grateful to M.E. Shirokov for an inspiration of this
work and many fruitful discussions. The work is partially
supported by intas grant Ref. Nr. 06-1000014-6077.

\end {document}